\documentclass[aps,pra,twocolumn,groupedaddress,showpacs]{revtex4}

\usepackage{graphicx}
\usepackage{amsmath}

\begin{document}


\title{Quadrupole-scissors modes and nonlinear mode coupling 
in trapped two-component Bose-Einstein condensates}


\author{Kenichi Kasamatsu$^1$}
\author{Makoto Tsubota$^1$}
\author{Masahito Ueda$^2$}

\affiliation{
$^1$Department of Physics,
Osaka City University, Sumiyoshi-Ku, Osaka 558-8585, Japan 
\\
$^2$Department of Physics, Tokyo Institute of Technology,  
Meguro-ku, Tokyo 152-8551, Japan}


\date{\today}

\begin{abstract}
We theoretically investigate quadrupolar collective excitations in two-component Bose-Einstein condensates and their nonlinear dynamics associated with harmonic generation and mode coupling. 
Under the Thomas-Fermi approximation and the quadratic polynomial ansatz for density fluctuations, the linear analysis of the superfluid hydrodynamic equations predicts excitation frequencies of three normal modes constituted from monopole and quadrupole oscillations, and those of three scissors modes. 
We obtain analytically the resonance conditions for the second harmonic generation in terms of the trap aspect ratio and the strength of intercomponent interaction. 
The numerical simulation of the coupled Gross-Pitaevskii equations vindicates the validity of the analytical results and reveals the dynamics of the second harmonic generation and nonlinear mode coupling that lead to nonlinear oscillations of the condensate with damping and recurrence reminiscent of the Fermi-Pasta-Ulam problem. 
\end{abstract}


\maketitle

\section{INTRODUCTION}
The study of collective excitations gives us insights into the fundamental properties of many-body systems. 
The experimental observations of collective excitations in alkali-atomic Bose-Einstein condensates (BECs) \cite{Jin,Mewes} have stimulated theoretical investigations in this context \cite{BECrev}. 
When the amplitude of a collective mode is increased, the dynamics of a BEC enters a nonlinear regime affected by the interatomic interaction, giving rise to a significant frequency shift and mode coupling in the collective excitations. 
Compared with a single-component BEC, even richer collective dynamics have been predicted to occur in two-component BECs due to the intercomponent interaction \cite{Busch,Graham,Ersy,Pu,Gordon,Sinatra}. 
A mixture of BECs can be produced experimentally by simultaneously trapping atoms in different hyperfine states \cite{Hall,Maddaloni,Stenger} or of different species \cite{Modugno}. 
Collective excitations and effects of the intercomponent interaction have been investigated experimentally by exciting a dipole mode, i.e., a center-of-mass oscillation, of two-component BECs \cite{Hall,Maddaloni,Modugno}. 
The dynamics of the center-of-mass oscillation was studied numerically by Sinatra {\it et al.} \cite{Sinatra}; they found two dynamic regimes of a periodic oscillation and a damped oscillation with strong nonlinear mixing of excited modes. 
The present paper investigates excitations of quadrupole and scissors modes in two-component BECs with a particular emphasis on the nonlinear dynamics due to the mode coupling between these modes. 

The frequencies of collective modes are accurately described by the linear analysis of the Gross-Pitaevskii (GP) equation. 
In the Thomas-Fermi (TF) limit, which applies to condensates with a large number 
of atoms, the GP equation reduces to a hydrodynamic equation, whose analytical solution for the excitation frequencies can be obtained by expressing the density fluctuations in terms of polynomials of degree $n$ in the Cartesian coordinates $x$, $y$ and $z$ \cite{BECrev}. 
The polynomial of $n=1$, e.g., $\delta n^{(1)} = \sum_{i=1}^{3} p_{i} x_{i}$, where $(x_{1},x_{2},x_{3})=(x,y,z)$, describes three dipole modes. 
The dipole modes have the frequencies of the harmonic trapping potential, independent of the interatomic interaction due to Kohn's theorem. 
The polynomial of $n=2$, e.g., $\delta n^{(2)} = \sum p_{ij} x_{i} x_{j}$, gives six normal modes. 
In an axisymmetric harmonic potential, linear combinations of the diagonal components $p_{xx}$, $p_{yy}$ and $p_{zz}$ describe three normal modes: one $m=2$ mode and two radial-breathing modes with $m=0$, where $m$ is the projected angular momentum on the symmetry axis; we shall refer to these modes as {\it diagonal quadrupole modes}. 
The remaining three normal modes are associated with the off-diagonal components $p_{xy}$, $p_{yz}$ and $p_{zx}$; we shall refer to them as {\it scissors modes}  \cite{Odelin}, by which superfluidity of a BEC has been demonstrated \cite{Marago}. 
The nonlinear term in the GP equation couples these modes more strongly for larger amplitudes of the excitation. 
The nonlinear mode coupling is also sensitive to the trap geometry because the excitation frequency depends on the trapping frequencies. 
Hechenblaikner {\it et al.} observed the second harmonic generation and mode coupling between the two $m=0$ radial-breathing modes in a single-component BEC by varying the aspect ratio of a trapping potential \cite{Hechenblaikner} and obtained the results consistent with the theoretical prediction \cite{Dalfovo,Hechenblaikner2}. The mode couplings associated with the scissors modes have also been studied experimentally \cite{Hodby} and theoretically \cite{Hechenblaikner2,Kawaja}. 

While the collective excitations in a single-component BEC have been well understood, much remains to be investigated for the collective excitations of two-component BECs. 
In this paper, we first focus on the quadrupole and scissors modes of two-component BECs, and discuss their mode couplings by numerical simulations of the GP equations. 
While collective excitations of two-component BEC systems have been studied both analytically \cite{Busch,Graham} and numerically \cite{Ersy,Pu,Gordon}, no special attention has been paid to quadrupole and scissors modes. 
From a theoretical point of view, it is not easy to predict analytically the collective excitation frequencies of two-component BECs in a trapping potential because of the complex ground-state structure \cite{Ho,Ohberg,Timmermans,Trippenbach,Riboli}. 
The assumption that two components overlap in the ground state allows us to use the polynomial ansatz for density fluctuations within the TF approximation and to analytically derive the frequencies of the collective modes. 
The quadratic polynomial ansatz gives us a clear understanding on three diagonal quadrupole modes and three off-diagonal scissors modes. The spectra of these modes are shown to bifurcate into the in-phase and out-of-phase oscillations due to the intercomponent interaction. 
Out numerical simulation of the coupled GP equation vindicates the validity of the analytical results. 
We obtain the resonance conditions of the second harmonic generation, in which two quanta of the fundamental mode are converted into one quantum of the second harmonic, depending on the trap aspect ratio and the strength of intercomponent interaction. 
Our numerical simulation clarifies the dynamics of the second harmonic generation and nonlinear mode coupling.

This paper is organized as follows. 
Section \ref{formulation} describes the basic equations of the system and briefly reviews the equilibrium properties of the system in the TF approximation. 
Under the assumption of the miscibility of two-component BECs and the quadratic polynomials for the density fluctuations, the frequencies of the diagonal quadrupole modes and the scissors modes are found in Sec. \ref{frequency}. 
In Sec. \ref{resoesti}, the resonance conditions of second harmonic generation of the normal modes are obtained. 
The nonlinear dynamics associated with the mode couplings is numerically studied in Sec. \ref{numeridis}. 
Section \ref{concl} summarizes and concludes this paper. 

\section{BASIC EQUATIONS AND EQUILIBRIUM STATE} \label{formulation}
We consider two-component BECs with the atomic mass $m_{i}$ in the hyperfine state $| F_{i},M_{Fi} \rangle$. 
The condensates are trapped by harmonic potentials which are described in terms of the magnetic field $|B({\bf r})|$ as \cite{Riboli}
\begin{eqnarray}
V_{i}({\bf r}) = \mu_{B} g_{i} M_{F_{i}} |B({\bf r})| \nonumber \\ 
\simeq \mu_{B} g_{i} M_{F_{i}} \biggl\{ B_{0} + \frac{1}{2} (K_{x}x^{2}+K_{y}y^{2}+K_{z}z^{2}) \biggr\} \nonumber \\
= V_{i}^{0} + \frac{1}{2} m_{i} ( \omega_{ix}^{2}  x^{2} + \omega_{iy}^{2}  y^{2} + \omega_{iz}^{2}  z^{2} ),
\hspace{2mm}  i=1,2, 
\end{eqnarray}
where $\mu_{B}$ is the Bohr magneton, $g_{i}$ the $g$-factor of the $i$-th component, and $\omega_{ik}$ ($k=x,y,z$) the trapping frequency satisfying the relation 
\begin{equation}
m_{i}\omega_{ik}^{2} = \mu_{B} g_{i} M_{F_{i}} K_{k}. 
\label{massfre}
\end{equation} 
In this paper, we do not consider the relative displacement of the potential minima caused by the constant term $V_{i}^{0}=\mu_{B} g_{i} M_{F_{i}} B_{0}$ by assuming $g_{1} M_{F1} = g_{2} M_{F2}$ \cite{tyuu}. 
We also neglect the displacement of the system due to gravitation \cite{Hall}.
The dynamics of two-component BECs are described by the coupled GP equations for the condensate wave functions $\psi_{1}$ and $\psi_{2}$,
\begin{subequations}
\begin{eqnarray}
i\hbar \frac{\partial \psi_{1}}{\partial t} = \biggl( -\frac{\hbar^{2} \nabla^{2}}{2m_{1}} +V_{1} + u_{1} |\psi_{1}|^{2} + u_{12} |\psi_{2}|^{2} \biggr) \psi_{1}, \\
i\hbar \frac{\partial \psi_{2}}{\partial t} = \biggl( -\frac{\hbar^{2} \nabla^{2}}{2m_{2}} +V_{2} + u_{2} |\psi_{2}|^{2} + u_{12} |\psi_{1}|^{2} \biggr) \psi_{2}. 
\end{eqnarray} \label{2tgpe}
\end{subequations}
Here the intracomponent interactions ($u_{1}$ and $u_{2}$) and the intercomponent one ($u_{12}$) are expressed in terms of the corresponding s-wave scattering lengths  $a_{1}$, $a_{2}$ and $a_{12}$ as $u_{i}=4\pi\hbar^{2}a_{i}/m_{i}$ ($i=1,2$) and $u_{12} = 2\pi\hbar^{2}a_{12}/m_{12}$ with $m_{12} \equiv m_{1}m_{2}/(m_{1}+m_{2})$ being the reduced mass. 
In this work, we will discuss the case of positive scattering lengths. 
By writing the condensate wave function as $\psi_{i} ({\bf r},t) = \sqrt{n_{i} ({\bf r},t)} e^{i \theta_{i} ({\bf r},t)}$, the time-dependent equations for the condensate density $n_{i}({\bf r},t)$ and the velocity field ${\bf v}_{i}=(\hbar/m_{i})\nabla \theta_{i}$ are given from Eqs. (\ref{2tgpe}a) and (\ref{2tgpe}b) as
\begin{subequations}
\begin{eqnarray}
\frac{\partial n_{1}}{\partial t} = - \nabla \cdot (n_{1} {\bf v}_{1}), \\
m_{1} \frac{\partial {\bf v}_{1}}{\partial t} = - \nabla \biggl( \frac{\hbar^{2}}{2 m_{1}} \frac{\nabla^{2} \sqrt{n_{1}}}{\sqrt{n_{1}}} + V_{1} - \mu_{1} \nonumber \\
+ u_{1} n_{1} + u_{12} n_{2} + \frac{1}{2} m_{1} v_{1}^{2} \biggr),  \\
\frac{\partial n_{2}}{\partial t} = - \nabla \cdot (n_{2} {\bf v}_{2}),  \\
m_{2} \frac{\partial {\bf v}_{2}}{\partial t} = - \nabla \biggl( \frac{\hbar^{2}}{2 m_{2}} \frac{\nabla^{2} \sqrt{n_{2}}}{\sqrt{n_{2}}} + V_{2} - \mu_{2} \nonumber \\
+ u_{2} n_{2} + u_{12} n_{1} + \frac{1}{2} m_{2}v_{2}^{2} \biggr). 
\end{eqnarray}
\label{hydroeq1}
\end{subequations}
The chemical potential $\mu_{i}$ is determined from the normalization condition $N_{i} = \int d {\bf r} n_{i}$ where $N_{i}$ is the number of particle of the $i$-th component. 

The frequencies of collective modes are calculated by linearizing Eqs. (\ref{hydroeq1}) around the equilibrium solutions as $n_{i}=n_{i}^{(0)} + \delta n_{i}$ and ${\bf v}_{i}= \delta {\bf v}_{i}$. 
The equilibrium solutions of the coupled GP equations show a variety of stable structures \cite{Ho,Ohberg,Timmermans,Trippenbach,Riboli}. 
When the intercomponent interaction is strongly repulsive, the two components separate spatially; the condition of the phase separation for trapped two-component BECs can be derived simply with the TF approximation \cite{Ho,Trippenbach,Riboli}, in which the quantum pressure terms $(\hbar^{2}/2m_{i}\sqrt{n_{i}}) \nabla^{2} \sqrt{n_{i}}$ in Eqs. (\ref{hydroeq1}b) and (\ref{hydroeq1}d) are neglected. 
In the spatial region where two components overlap ($n_{1}^{(0)} \neq 0$ and $n_{2}^{(0)} \neq 0$),  the equilibrium density distributions are given by
\begin{subequations}
\begin{eqnarray}
n_{1}^{(0)} &=& \frac{u_{2} \mu_{1} -u_{12} \mu_{2}-u_{2} V_{1} + u_{12} V_{2}}{u_{1}u_{2} - u_{12}^{2}}, \nonumber \\
&=& \frac{1}{u_{1} \Gamma_{12}} \biggl\{ \mu_{1} - \frac{u_{12}}{u_{2}} \mu_{2} - \Gamma_{2} V_{1}({\bf r}) \biggr\},  \\
n_{2}^{(0)} &=& \frac{u_{1} \mu_{2} -u_{12} \mu_{1}-u_{1} V_{2} + u_{12} V_{1}}{u_{1}u_{2} - u_{12}^{2}}  \nonumber \\
&=& \frac{1}{u_{2} \Gamma_{12}} \biggl\{ \mu_{2} - \frac{u_{12}}{u_{1}} \mu_{1} - \Gamma_{1} V_{2}({\bf r}) \biggr\}.
\end{eqnarray}
\label{2compTFden}
\end{subequations}
Here we defined the following dimensionless parameters,
\begin{eqnarray}
\Gamma_{12} \equiv 1-\frac{m_{1}+m_{2}}{4m_{12}} \frac{a_{12}^{2}}{a_{1}a_{2}}, \\
\Gamma_{1} \equiv 1-\frac{m_{1}}{2m_{12}} \frac{a_{12}}{a_{1}}, \\
\Gamma_{2} \equiv 1-\frac{m_{2}}{2m_{12}} \frac{a_{12}}{a_{2}}. 
\label{condito}
\end{eqnarray}
In order for the two components to overlap, the right-hand sides of Eqs. (\ref{2compTFden}a) and (\ref{2compTFden}b) must be positive. The overlapping region of the two components is delimited by the boundary defined as $\mu_{i} - u_{12} \mu_{j} / u_{j}  - \Gamma_{j} V_{i}({\bf r})=0$ \cite{Riboli}. 
In addition, the parameter $\Gamma_{12}$ must be positive, for if $\Gamma_{12}<0$, i.e., $u_{1}u_{2}-u_{12}^{2}<0$, there is no overlapping region, which implies the phase separation of the two components. 
The phase-separated solutions will not be considered in this work \cite{Svidzinsky}. 
For $\Gamma_{12}>0$, there are two possible overlapping density profiles depending on the signs of the parameters $\Gamma_{1}$ and $\Gamma_{2}$. 
These signs determine the curvature of the density profile at the center of the trapping potential because $\Gamma_{1}$ and $\Gamma_{2}$ are the coefficients of the $x^{2}$, $y^{2}$ and $z^{2}$ terms in Eqs. (\ref{2compTFden}a) and (\ref{2compTFden}b).
For $\Gamma_{1}>0$ and $\Gamma_{2}>0$, both density profiles are written by inverted parabolas in the overlapping region. 
For $\Gamma_{1}>0$ and $\Gamma_{2}<0$ ($\Gamma_{1}<0$ and $\Gamma_{2}>0$), one density profile is an inverted parabola, but the other is a usual parabola in the overlapping region. 
In the peripheral region in which the density of one component vanishes, the other component follows the single-component TF profile $n_{i}^{(0)}=(\mu_{i}-V_{i})/u_{i}$. 
This profile is connected to the overlapping density profile of Eq. (\ref{2compTFden}) by adjusting the chemical potential $\mu_{i}$ ($i=1$ or $2$) under the normalization condition $\int d {\bf r} n_{i}^{(0)} = N_{i}$ \cite{Ho}. 

\section{FREQUENCIES OF THE LOW-LYING COLLECTIVE MODES} \label{frequency}
In this section we discuss the excitation frequencies of the diagonal quadrupole modes and the scissors modes of the two-component BECs by solving Eqs. (\ref{hydroeq1}). 
In the following analysis, we assume that the two components overlap completely, use Eqs. (\ref{2compTFden}a) and (\ref{2compTFden}b) with inverted parabolas ($\Gamma_{1}>0$ and $\Gamma_{2}>0$) as the equilibrium solution $n_{i}^{(0)}$, and apply no boundary conditions to the solutions of Eqs. (\ref{hydroeq1}a) and (\ref{hydroeq1}b). 
This assumption may be effective when the TF radii 
\begin{equation}
R_{1k}^{\rm TF} = \sqrt{ \frac{\mu_{1}-u_{12} \mu_{2} / u_{1}}{\Gamma_{2} m_{1} \omega_{1k}^{2}/2} }, \hspace{2mm}
R_{2k}^{\rm TF} = \sqrt{ \frac{\mu_{2}-u_{12} \mu_{1} / u_{2}}{\Gamma_{1} m_{2} \omega_{2k}^{2}/2} }
\end{equation}
($k=x,y,z$) of the two overlapping condensates are almost equal: 
\begin{equation}
\frac{R_{2k}^{\rm TF}}{R_{1k}^{\rm TF}} = \biggl( \frac{u_{2}}{u_{1}} 
\frac{\Gamma_{2}}{\Gamma_{1}} \frac{N_{2}}{N_{1}} \biggr)^{1/5} \simeq 1.
\label{conditTF}
\end{equation}
In addition, the amplitude of the fluctuation should be small for the assumption to be valid. 
As shown later, the predicted frequencies of collective modes are independent of the particle number under the TF approximation. Therefore, for any $m_{i}$ and $a_{i}$ the condition (\ref{conditTF}) can be met by adjusting the ratio of the particle number $N_{2}/N_{1}$. 

The frequencies of collective modes are obtained by linearizing Eqs. (\ref{hydroeq1}a) and (\ref{hydroeq1}b) around the equilibrium solutions as $n_{i}=n_{i}^{(0)} + \delta n_{i}$ and ${\bf v}_{i}= \delta {\bf v}_{i}$. Here we neglect the quantum pressure term in Eqs (\ref{hydroeq1}). 
This results from the assumption that the spatial variations of the density are smooth not only in the ground state but also in the low-lying excited states. 
The linearized equations then read 
\begin{subequations}
\begin{eqnarray}
\frac{\partial^{2} \delta n_{1}}{\partial t^{2}} 
= \frac{1}{m_{1}} \nabla \cdot \{ n_{1}^{(0)} \nabla (u_{1} \delta n_{1} + u_{12} \delta n_{2}) \}, \\
\frac{\partial^{2} \delta n_{2}}{\partial t^{2}} 
= \frac{1}{m_{2}} \nabla \cdot \{ n_{2}^{(0)} \nabla (u_{2} \delta n_{2} + u_{12} \delta n_{1}) \}. 
\end{eqnarray}\label{linhydroeq}
\end{subequations}
The analytic solutions to these equations are obtained by expressing the density fluctuations in terms of quadratic polynomials \cite{BECrev}: 
\begin{subequations}
\begin{eqnarray}
\delta n_{1}({\bf r},t) = p_{0}(t) - p_{x}(t) x^{2} - p_{y}(t) y^{2} -p_{z}(t) z^{2} \nonumber \\
 - p_{xy}(t) x y - p_{yz}(t) y z -p_{zx}(t) z x,  \\
 \delta n_{2}({\bf r},t) = q_{0}(t) - q_{x}(t) x^{2} - q_{y}(t) y^{2} -q_{z}(t) z^{2} \nonumber \\
 - q_{xy}(t) x y - q_{yz}(t) y z -q_{zx}(t) z x. 
\end{eqnarray}
\label{fluctu12}
\end{subequations}
The diagonal quadrupole modes are described by linear combinations of the diagonal components $p_{x,y,z}$ and $q_{x,y,z}$, while the off-diagonal components $p_{xy,yz,zx}$ and $q_{xy,yz,zx}$ determine the scissors modes \cite{Odelin,Kawaja,Hechenblaikner2}. 
The time dependences of $p_{0}$ and $q_{0}$ are determined by the conditions $\int d {\bf r} \delta n_{i} = 0$, which express the particle-number conservation of each component. 
The equilibrium states $n_{i}^{(0)}$ are given by Eqs. (\ref{2compTFden}a) (\ref{2compTFden}b). 
It follows then from Eqs. (\ref{linhydroeq}a) and (\ref{linhydroeq}b) that the time-dependent variables ${\bf X}=(p_{x}, q_{x}, p_{y}, q_{y}, p_{z}, q_{z}, p_{xy}, q_{xy}, p_{yz}, q_{yz}, p_{zx}, q_{zx})^{T}$ obey
\begin{eqnarray}
\frac{d^{2}}{dt^{2}} {\bf X} = - {\bf M} \cdot {\bf X} = -
\left(
\begin{array}{cc}
{\bf M}_{\rm Q} & 0  \\
0 & {\bf M}_{\rm S} \\
\end{array}
\right) \cdot  {\bf X}. 
\label{linearmat}
\end{eqnarray} 
In the linear limit, the diagonal quadrupole modes and the off-diagonal scissors modes are completely decoupled \cite{Kawaja,Hechenblaikner2}. 
The matrix ${\bf M}_{\rm Q}$ is too large to be written down here, but ${\bf M}_{\rm S}$ for the scissors modes is simply given as 
\begin{eqnarray}
{\bf M}_{\rm S} = 
\left(
\begin{array}{ccc}
{\bf M}_{xy} & 0 & 0 \\
0 & {\bf M}_{yz} & 0 \\
0 & 0 & {\bf M}_{zx}
\end{array}
\right) \label{scimatform}
\end{eqnarray}
with
\begin{eqnarray}
{\bf M}_{jk} = 
\left(
\begin{array}{cc}
\frac{\Gamma_{2}}{\Gamma_{12}} (\omega_{1j}^{2} + \omega_{1k}^{2}) & 
\frac{\Gamma_{2}}{\Gamma_{12}} \frac{u_{12}}{u_{1}} (\omega_{1j}^{2} + \omega_{1k}^{2}) \\
\frac{\Gamma_{1}}{\Gamma_{12}} \frac{u_{12}}{u_{2}} (\omega_{2j}^{2} + \omega_{2k}^{2})  & 
\frac{\Gamma_{1}}{\Gamma_{12}} (\omega_{2j}^{2} + \omega_{2k}^{2}) 
\end{array}
\right).
\end{eqnarray}
Since the three scissors modes are also decoupled as seen from Eq. (\ref{scimatform}), we can treat the $xy$ scissors mode separately. 
The frequency $\Omega_{xy}$ is easily calculated from ${\bf M}_{xy}$ by substituting $p_{xy}=T_{1} e^{i \Omega_{xy} t}$ and $q_{xy}=T_{2} e^{i \Omega_{xy} t}$ into Eq. (\ref{linearmat}) and solving the resulting secular equation with the result 
\begin{eqnarray}
\Omega_{xy}^{2} = \frac{1}{2 \Gamma_{12}} 
\biggl[ \Gamma_{2} (\omega_{1x}^{2} + \omega_{1y}^{2})
+ \Gamma_{1} (\omega_{2x}^{2} + \omega_{2y}^{2})  \nonumber \\
\pm \biggl\{ 4 \frac{u_{12}^{2}}{u_{1}u_{2}} \Gamma_{1} \Gamma_{2} 
(\omega_{1x}^{2} + \omega_{1y}^{2}) (\omega_{2x}^{2} + \omega_{2y}^{2}) \nonumber \\
+\left[ \Gamma_{2} (\omega_{1x}^{2} + \omega_{1y}^{2}) 
-\Gamma_{1} (\omega_{2x}^{2} + \omega_{2y}^{2}) \right]^{2} \biggr\}^{1/2} \biggr].  
\label{scissorsfreq}  
\end{eqnarray}
For the case of equal atomic mass $m_{1}=m_{2}$  (hence $\omega_{1k}=\omega_{2k}=\omega_{k}$ by Eq. (\ref{massfre})), Eq. (\ref{scissorsfreq}) is simplified to 
\begin{subequations}
\begin{eqnarray}
\Omega_{xy+} &=& \sqrt{\omega_{x}^{2} + \omega_{y}^{2}},  \\
\Omega_{xy-} &=& \sqrt{\omega_{x}^{2} + \omega_{y}^{2}} \sqrt{\delta a} 
\end{eqnarray}
\label{phasesimple}
\end{subequations}
with
\begin{eqnarray}
\delta a = \frac{(a_{1}-a_{12})(a_{2}-a_{12})}{a_{1}a_{2}-a_{12}^{2}}. 
\label{smatome}
\end{eqnarray}
When the two components overlap completely, we have $0 < \delta a \leq 1$, so that $\Omega_{xy-} \leq \Omega_{xy+}$. 
Thus, the frequencies of the scissors modes split into two branches. 
The upper frequency $\Omega_{xy+}$ correspond to the in-phase oscillation which keeps the principal axes of the two components in line. For $m_{1}=m_{2}$, the frequency of the in-phase mode is independent of the intercomponent interactions because of Kohn's theorem \cite{BECrev}. 
The lower frequency $\Omega_{xy-}$ corresponds to the out-of-phase motion of the two components in which they oscillate with the $180^{\circ}$ phase difference. 
The frequency of the out-of-phase mode depends strongly on the interatomic interaction; the measurement of both frequencies of the in-phase and out-of-phase modes thus allows us to determine $\delta a$ from Eqs. (\ref{phasesimple}a) and (\ref{phasesimple}b). 

The frequencies of the diagonal quadrupole modes are also calculated by solving the corresponding secular equation of ${\bf M}_{\rm Q}$. 
For $m_{1}=m_{2}$ and the axisymmetry of a trapping potential $\omega_{x}=\omega_{y}/\lambda=\omega_{z}$, the secular equation for ${\bf M}_{\rm Q}$ yields 
\begin{subequations}
\begin{eqnarray}
\frac{\Omega_{+}^{2}}{\omega_{x}^{2}} &=& \biggl\{ 
\begin{array}{l}
2  \\ 
2 + \frac{3}{2}\lambda^{2} 
\mp \frac{1}{2} \sqrt{9 \lambda^{4} - 16 \lambda^{2} +16}
\end{array}, \\
\frac{\Omega_{-}^{2}}{\omega_{x}^{2}} &=& \frac{\Omega_{+}^{2}}{\omega_{x}^{2}} \delta a. 
\end{eqnarray} \label{quadfreq}
\end{subequations}

First, we quote the excitation of a single component condensate, i.e., $a_{12}=0$ ($\delta a = 1$). 
When $\lambda=\sqrt{8}$ and $a_{12}=0$, for example, Eqs. (\ref{quadfreq}) give 
$\Omega/\omega_{x}=\sqrt{2}$, $1.79$ and $4.97$ \cite{Hodby}, which corresponds to a quadrupole-type oscillation in the $x$-$z$ plane with $m=2$, 
an in-phase oscillation in the $x$ and $z$ directions but out-of-phase in the  $y$ direction, i.e., the $m=0$ low-lying mode, and an in-phase compressional mode in all directions i.e., the $m=0$ high-lying or breathing mode, respectively. 
Three scissors-mode frequencies are given by $\Omega_{zx}/\omega_{x}=\sqrt{2}$ and $\Omega_{xy}/\omega_{x}=\Omega_{yz}/\omega_{x}=3$; the $xz$-scissors mode is degenerate with the $m=2$ quadrupole mode \cite{Hodby}. 

\begin{figure}[tp]
\includegraphics[height=0.46\textheight]{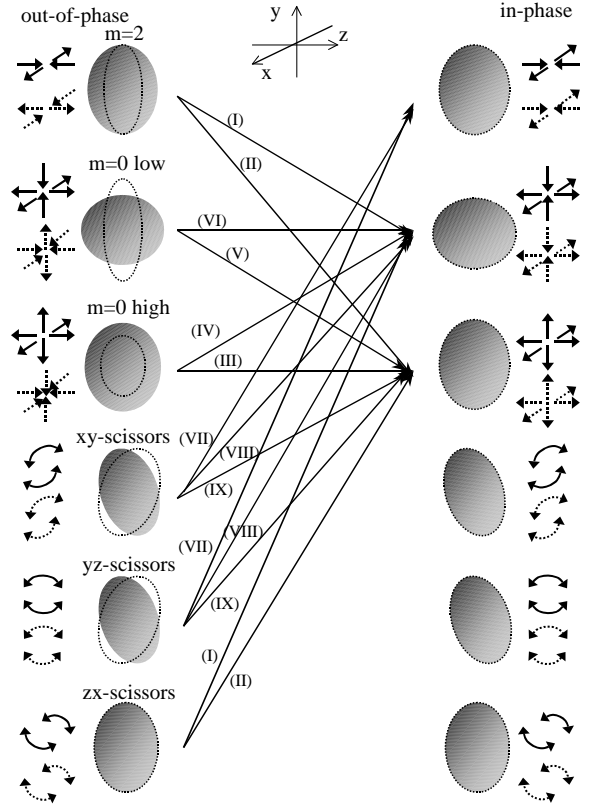}
\caption{Schematic representation of the quadrupole and scissors modes 
of two-component BECs. Oscillations of component 1 are indicated by solid arrows 
and those of component 2 by dashed arrows. 
Mode coupling processes in Table \ref{tab:table2} are also shown. }
\label{modepict}
\end{figure}
\begin{figure}[tp]
\includegraphics[height=0.48\textheight]{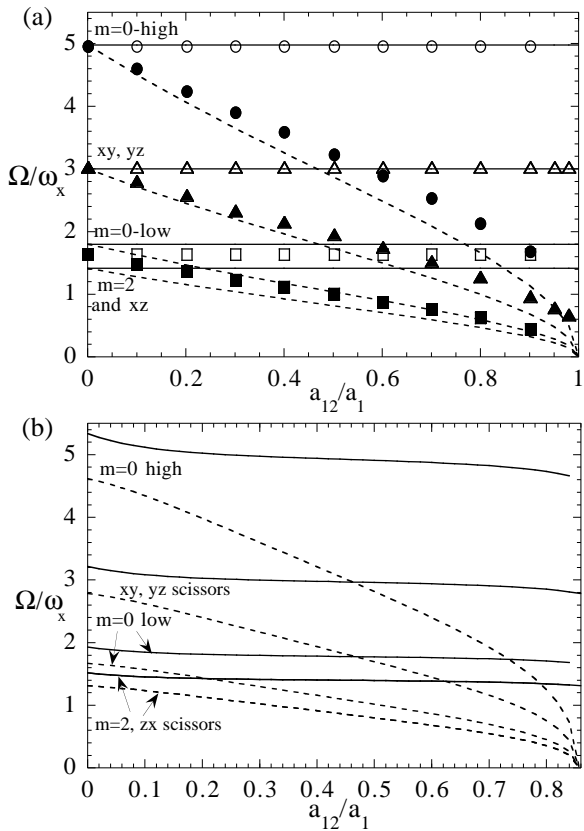}
\caption{$a_{12}/a_{1}$ dependences of the twelve collective normal modes obtained from Eq. (\ref{linearmat}) for the parameter $a_{1}=a_{2}$, $\lambda=\sqrt{8}$, and (a) $m_{2}=m_{1}$ and (b) $m_{2}=0.75m_{1}$. 
Each mode splits into the in-phase mode (thin curves) and the out-of-phase mode (dashed curves) for $a_{12} \neq 0$. In (a), the numerically obtained frequencies are plotted by open symbols (for in-phase) and filled symbols (for out-of -phase).
As explained in Sec. \ref{numeridis}, the reduction to a two-dimensional system changes the frequencies of the $m=0$ low- and high-lying modes; for $\lambda=\sqrt{8}$ their values become small from those of Eqs. (\ref{quadfreq}) by 10 \% for the low-lying mode and 1\% for the high-lying mode.}
\label{ep=0m2=m1}
\end{figure}

For $a_{12} \neq 0$ the frequencies of all modes split into two branches as mentioned above; the corresponding modes are graphically represented in Fig. \ref{modepict} . 
In Fig. \ref{ep=0m2=m1}(a), we show the frequencies of twelve collective modes as functions of $a_{12}/a_{1}$ for the condensates with $a_{1}=a_{2}$.
The frequencies of the out-of-phase modes decrease monotonically with $a_{12}$, because the stronger mutual repulsion favors the phase separation and makes these modes softer. 
The frequencies vanish at $a_{12}/a_{1}=1$, at which the linear analysis around $n_{i}^{(0)}$ of Eqs. (\ref{2compTFden}a) and (\ref{2compTFden}b) with an inverted parabola breaks down. 

For $m_{1} \neq m_{2}$ the excitation frequencies of all modes separate even at $a_{12}=0$, because the trapping frequency of one component is different from that of the other as seen in Eq. (\ref{massfre}). 
The frequencies for $m_{2}=0.75m_{1}$ are shown in Fig. \ref{ep=0m2=m1}(b). 
In this case, the frequencies corresponding to the in-phase motion decrease gradually with $a_{12}$, but the qualitative behavior is nearly the same as that in Fig. \ref{ep=0m2=m1}(a). 
In general, the frequency of the out-of-phase modes vanishes when
\begin{equation}
\Gamma_{1} \Gamma_{2} = 
\biggl( 1-\frac{m_{1}}{2m_{12}} \frac{a_{12}}{a_{1}}\biggr) 
\biggl( 1-\frac{m_{2}}{2m_{12}} \frac{a_{12}}{a_{2}}\biggr) = 0, 
\end{equation}
which gives, e.g., $a_{12}/a_{1}=0.856$ for $m_{2}=0.75m_{1}$. 
These results are consistent with the numerical study of the excitation frequency \cite{Pu}, where the frequency of the out-of-phase mode also decreases with $a_{12}$ and the ground state undergoes phase separation eventually. 

\section{HARMONIC GENERATION AND MODE COUPLNG AMONG QUADRUPOLAR MODES}\label{numesci}
In the preceding section, we obtained the frequencies of the diagonal quadrupole modes and the scissors modes with the in-phase or out-of-phase motions of two components. 
Although these modes are decoupled in the linear limit, the growth of the excitation amplitude makes the nonlinear coupling among these modes stronger, casting the dynamics into a nonlinear regime. 
The mode coupling also depends strongly on the geometry of a trapping potential \cite{Dalfovo,Hechenblaikner,Kawaja,Hechenblaikner2,Hodby}. 
In this section, we will show the nonlinear dynamics caused by the harmonic generation and the mode coupling among the quadrupole excitations by numerically solving the coupled GP equations (\ref{2tgpe}). 
The nonlinear term in the GP equation is proportional to $|\psi|^{2}$, being able to generate an excitation with a frequency of the integral multiple of that of an initially excited mode. 
This fact allows the spontaneous up-conversion process, while the down-conversion cannot be explained within the framework of GP equation \cite{Kawaja,Hechenblaikner2,Hodby}. 
In this work, we discuss the up-conversion process of the harmonic generation and confine ourselves to the second-harmonic process, 
where the resonance of two modes occurs via annihilation (creation) of two quanta in a lower-energy mode and creation (annihilation) of one quantum in a higher-energy mode i.e., second harmonic generation at $\Omega_{\rm high}=2\Omega_{\rm low}$. 
Compared with the single-component case \cite{Dalfovo,Hechenblaikner,Kawaja,Hechenblaikner2,Hodby}, an even richer variety of mode couplings can occur because of the presence of the out-of-phase modes, caused by the intercomponent interaction. 

\subsection{Resonance condition}\label{resoesti}
Before discussing the numerical results, we evaluate the resonance condition between two quadrupole modes derived in the preceding section. In order to simplify the problem, we consider two-component condensates with the equal mass ($m_{1}=m_{2}$ and $\omega_{1k}=\omega_{2k}$), trapped in an axially symmetric potential $(\omega_{x}=\omega_{y}/\lambda=\omega_{z})$. 
Then, simple algebraic manipulations of Eqs. (\ref{phasesimple}) and (\ref{quadfreq}) yield some resonance conditions for the mode coupling. 

\subsubsection{Coupling between two diagonal quadrupole modes}
First, we consider the mode coupling between two diagonal quadupole modes. 
To take account of the nonlinear effect, we follow the method of Ref. \cite{Dalfovo}; the density and the velocity are assumed to take the forms $n_{1}({\bf r},t)=p_{0}(t) - p_{x}(t) x^{2} - p_{y}(t) y^{2} -p_{z}(t) z^{2}$ and ${\bf v}_{1}({\bf r},t)=\nabla \{ \alpha_{x}(t)x^{2}+\alpha_{y}(t)y^{2}+\alpha_{z}(t)z^{2} \}$, and similar forms for component 2 by replacing $p \rightarrow q$ and $\alpha \rightarrow \beta$, where we neglect the off-diagonal contribution. 
The substitution of these terms into Eqs. (\ref{hydroeq1}) yields the equations of motion for the time-dependent variables $p,q,\alpha$ and $\beta$. As in Sec. \ref{frequency}, we impose no boundary condition on these variables. 
These equations are simplified by introducing the new variables $a_{k}$ and $b_{k}$ ($k=x,y,z$) defined by \cite{Dalfovo}
\begin{equation}
p_{k}=\frac{\Gamma_{2} m_{1} \omega_{1k}^{2}}{2 u_{1} \Gamma_{12} a_{x} a_{y} a_{z} a_{k}^{2}}, \hspace{3mm}
q_{k}=\frac{\Gamma_{1} m_{2} \omega_{2k}^{2}}{2 u_{2} \Gamma_{12} b_{x} b_{y} b_{z} b_{k}^{2}}. 
\label{hensuhenkan}
\end{equation}
Then we obtain $\alpha_{k} = \dot{a}_{k}/2a_{k}$, $\beta_{k} = \dot{b}_{k}/2b_{k}$, and 
\begin{eqnarray}
\ddot{a_{k}}= \omega_{1k}^{2} \biggl (- a_{k} + \frac{\Gamma_{2}}{\Gamma_{12}} 
\frac{1}{a_{x} a_{y} a_{z} a_{k}} +\frac{u_{12} \Gamma_{1}}{u_{2} \Gamma_{12}} 
\frac{a_{k}}{b_{x}b_{y}b_{z}b_{k}^{2}} \biggr ), \label{colleceq1} \\
\ddot{b_{k}}=\omega_{2k}^{2} \biggl (- b_{k} + \frac{\Gamma_{1}}{\Gamma_{12}} 
\frac{1}{b_{x} b_{y} b_{z} b_{k}} +\frac{u_{12} \Gamma_{2}}{u_{1} \Gamma_{12}} 
\frac{b_{k}}{a_{x}a_{y}a_{z}a_{k}^{2}} \biggr ). \label{colleceq2}
\end{eqnarray}
The linearization of Eqs. (\ref{colleceq1}) and (\ref{colleceq2}) yields the matrix ${\bf M}'_{\rm Q}$, 
whose eigenvalues are the same as those of ${\bf M}_{\rm Q}$ of Eq. (\ref{linearmat}). 
In order to clarify the mode coupling, we rewrite $a_{k}$ and $b_{k}$ by using the new basis obtained by the eigenvectors ${\bf e}_{i}$ of ${\bf M}'_{\rm Q}$ as ${\bf x}=\sum_{i=1}^{6} c_{i}(t) {\bf e}_{i}$, where ${\bf x}=(a_{x},a_{y},a_{z},b_{x},b_{y},b_{z})$. 
From Eqs. (\ref{colleceq1}) and (\ref{colleceq2}) one obtains the equations for the time-dependent variables $c_{i}(t)$; these equations are completely decoupled in the linear limit. 

In order to understand the process of second harmonic generation, we expand the equations of motion up to second order in $c_{i}(t)$, obtaining the form 
\begin{equation}
\ddot{c}_{i}= - \Omega_{i}^{2} c_{i} + \sum_{j,k} \gamma_{jk} c_{j} c_{k}
\end{equation} 
with the mode frequency $\Omega_{i}$. 
The second-harmonic process between the $i$- and $j$-modes occurs via the terms like $\gamma_{jj} c_{j}^{2}$ $(i \neq j)$ for $\Omega_{i} \simeq 2 \Omega_{j}$. 
For a single-component condensate ($a_{12}=0$) in an axially symmetric potential, one can easily show that at $\lambda=\sqrt{16/7}$ the second harmonic of the $m=2$ mode coincides with the $m=0$ high-lying mode, and at $\lambda=(5\sqrt{5} \pm \sqrt{29})/6\sqrt{2}$ the second harmonic of the $m=0$ low-lying mode coincides with the $m=0$ high-lying mode \cite{Dalfovo}. 
For two-component condensates with $m_{1}=m_{2}$, these conditions for $\lambda $ also hold for the mode couplings between in-phase modes. 

\begin{figure}[btp]
\includegraphics[height=0.46\textheight]{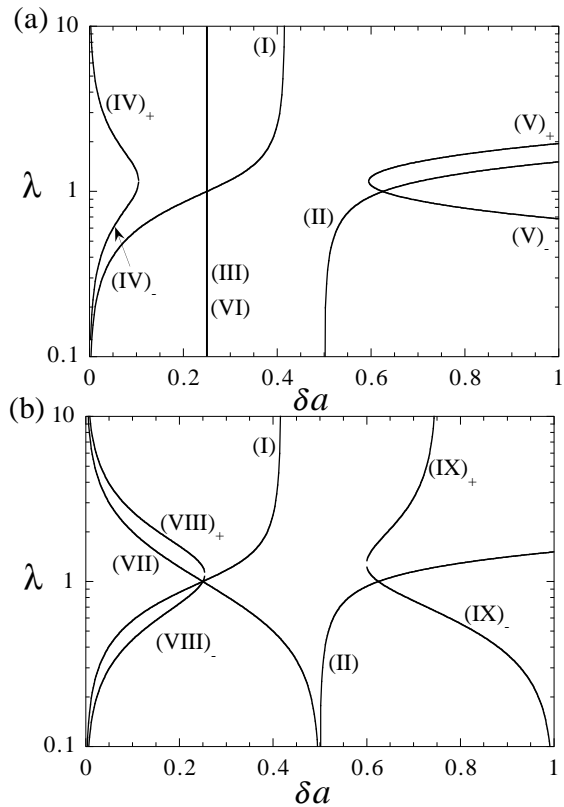}
\caption{Conditions of the second-harmonic resonance in the parameter space of $\lambda$  
and $\delta a = (a_{1}-a_{12}) (a_{2}-a_{12}) / (a_{11}a_{22}-a_{12}^{2}) $. 
The mode couplings between two diagonal quadrupole modes are shown in (a), and those between a diagonal quadrupole mode and a scissors mode in (b).}
\label{resonance}
\end{figure}

\begin{table*}
\caption{\label{tab:table2} List of all possible second-harmonic processes 
for two-component BECs with $m_{1}=m_{2}$ in axisymmetry potentials.  
IP and OP stands for ``in-phase" and ``out-of-phase", respectively.}
\begin{ruledtabular}
\begin{tabular}{cccc}
 & second-harmonic process & aspect ratio $\lambda=\omega_{y}/\omega_{x}$ & range of $\delta a$ \\
\hline
(I) & $m=2$ or $zx$-scissors, OP $\rightarrow$ $m=0$ - low, IP & 
$\lambda = 4 \sqrt{\frac{\delta a (1-2 \delta a)}{ 5-12 \delta a }}$ & 
0 $<$ $\delta a$ $<$ 5/12 \\
(II) & $m=2$ or $zx$-scissors, OP $\rightarrow$ $m=0$ - high, IP & 
$\lambda = 4 \sqrt{\frac{\delta a (1-2 \delta a)}{ 5-12 \delta a }}$ & 
1/2 $<$ $\delta a$ $<$ 1 \\
(III) & $m=0$-high, OP $\rightarrow$ $m=0$ - high, IP & 
$\lambda =  {\rm arbitrary}$ & $\delta a=0.25$ \\
(IV) & $m=0$ - high, OP $\rightarrow$ $m=0$ - low, IP & 
$\lambda = \frac{\sqrt{80 \delta a^{2} -8 \delta a + 5
\pm \sqrt{5} (1 + 4 \delta a) \sqrt{5 - 56 \delta a + 80 \delta a^{2}} }}{6 \sqrt{\delta a}} $ & 
0 $<$ $\delta a$ $<$  $\frac{7-2\sqrt{6}}{20}$ \\
(V) & $m=0$ - low, OP $\rightarrow$ $m=0$ - high, IP & 
$\lambda = \frac{\sqrt{80 \delta a^{2} -8 \delta a + 5
\pm \sqrt{5} (1 + 4 \delta a) \sqrt{5 - 56 \delta a + 80 \delta a^{2}} }}{6 \sqrt{\delta a}} $ & 
$\frac{7+2\sqrt{6}}{20}$ $<$ $\delta a$ $<$ 1 \\
(VI) & $m=0$-low, OP $\rightarrow$ $m=0$ - low, IP & 
$\lambda =  {\rm arbitrary}$ & $\delta a=0.25$ \\
(VII) & $xy,yz$-scissors, OP $\rightarrow$ $m=2$, IP & 
$\lambda =\sqrt{\frac{1-2 \delta a}{2 \delta a}}$ & 
$ 0 < \delta a < \frac{1}{2}$ \\
(VIII) & $xy,yz$-scissors, OP $\rightarrow$ $m=0$ - low, IP & 
$\lambda =\frac{1}{2} \sqrt{\frac{16 \delta a^{2} - 14 \delta a + 5 
\pm \sqrt{164 \delta a^{2} -140 \delta a + 25} }{\delta a (3 - 4 \delta a)}} $ & 
$ 0 < \delta a < \frac{5 (7-2\sqrt{2})}{82}$ \\
(IX) & $xy,yz$-scissors, OP $\rightarrow$ $m=0$ - high, IP & 
$\lambda =\frac{1}{2} \sqrt{\frac{16 \delta a^{2} - 14 \delta a + 5 
\pm \sqrt{164 \delta a^{2} -140 \delta a + 25} }{\delta a (3 - 4 \delta a)}} $ & 
$\frac{5 (7+2\sqrt{2})}{82} < \delta a < \biggl\{ 
\begin{array}{ll}
\frac{3}{4} & {\rm for} (+) \\ 
1 &  {\rm for} (-)  
\end{array} $
\end{tabular}
\end{ruledtabular}
\end{table*}
We summarize the other possible second-harmonic processes in Table \ref{tab:table2} and Fig. \ref{modepict}. 
The condition of the mode coupling is obtained by $\Omega_{i} = 2 \Omega_{j}$ in Eqs. (\ref{quadfreq}), being represented by the aspect ratio $\lambda$ and the interaction parameter $\delta a$ as shown in the middle column of Table \ref{tab:table2} and in Fig. \ref{resonance}(a). 
Since $\lambda>0$ and $0 < \delta a \leq 1$, depending on the relation between two frequencies, each coupling process occurs within a certain range of $\delta a$ as shown in the right column in Table \ref{tab:table2}. 
The processes (III) and (VI), which describes the coupling between the in-phase and out-of-phase modes with $m=0$ mode, occur for arbitrary $\lambda$ when $\delta a=0.25$. 
For condensates with equal mass there is no up-conversion process in which higher-lying modes are the $m=2$ modes and the out-of-phase $m=0$ modes, because the term $\gamma_{jj} c_{j}^{2}$ does not appear for these modes. (For different mass $m_{1} \neq m_{2}$, the out-of-phase $m=0$ modes can become higher-lying modes of the up-conversion process.)

\subsubsection{Coupling between a diagonal quadrupole mode and a scissors mode}
Now we discuss the mode coupling related with the off-diagonal scissors modes. 
When the off-diagonal contribution is included in the fluctuation, the analysis in the preceding subsection cannot be used because an appropriate transformation like Eq. (\ref{hensuhenkan}) does not exist. 
However, if we expand the condensate wave function as $\psi_{i}({\bf r},t) = e^{-i\mu_{i}t/\hbar}[\psi_{ig}({\bf r}) + u_{i}({\bf r}) e^{-i \Omega_{i} t} + v^{\ast}_{i}({\bf r}) e^{ i \Omega_{i} t}]$, the presence of the mode coupling is understood by the symmetry of the mode functions $u_{i}({\bf r})$ and $v_{i}({\bf r})$. 
The coupling coefficient for the second-harmonic process is written by the overlap integral of the ground state wave function and the mode functions as $\int d{\bf r} \psi_{ig}({\bf r}) u_{\rm low}^{\ast}({\bf r})^{2} u_{\rm high}({\bf r})$ \cite{Kawaja,Hechenblaikner2}. 
The scissors modes are characterized by the odd-parity function $u_{i}({\bf r}) \propto xy, yz, zx$. 
Thus, there is no coupling between a higher-lying scissors mode and a lower-lying diagonal quadrupole mode which has the even parity, because the overlap integral vanishes. 
However, there can be second-harmonic processes from a lower-lying scissors mode to a higher-lying diagonal quadrupole mode. 

The resonance condition for the $zx$-scissors mode is the same as that of the diagonal $m=2$ quadrupole mode, because they are degenerate in an axisymmetric potential. 
Thus, we focus on the mode coupling between the $xy$- and $yz$-scissors modes with the same frequency. 
For their in-phase mode, no resonance occurs for the diagonal in-phase or out-of-phase quadrupole modes, because the condition $\Omega_{i}=2\Omega_{j}$ cannot be satisfied in the case of an axisymmetric trap. 
On the other hand, there are couplings for their out-of-phase modes as shown in (VII), (VIII), and (IX) in Table \ref{tab:table2}, Fig. \ref{modepict} and Fig. \ref{resonance}(b). 

\subsubsection{Coupling between scissors modes}
A similar symmetry argument shows that there is no direct coupling via a second-harmonic process between two scissors modes, because the overlap integral of the second harmonic generation is always zero. 
Actually, if we neglect the diagonal modes, substitutions of the form $\delta n_{1} = p_{0}(t) - p_{xy}(t) xy - p_{yz}(t) yz -p_{zx}(t) zx$ and $\delta \alpha_{1} = \nabla \{- \alpha_{xy}(t) xy - \alpha_{yz}(t) yz -\alpha_{zx}(t) zx \}$ etc. into Eqs. (\ref{hydroeq1}) yields 
\begin{subequations}
\begin{eqnarray}
\dot{p}_{xy}=-2(p^{(0)}_{x}+p^{(0)}_{y}) \alpha_{xy} - p_{zx} \alpha_{yz} - p_{yz} \alpha_{zx},  \\
\dot{p}_{yz}=-2(p^{(0)}_{y}+p^{(0)}_{z}) \alpha_{yz} - p_{xy} \alpha_{zx} - p_{zx} \alpha_{xy},  \\
\dot{p}_{zx}=-2(p^{(0)}_{z}+p^{(0)}_{x}) \alpha_{zx} - p_{yz} \alpha_{xy} - p_{xy} \alpha_{yz},  \\
\dot{\alpha}_{xy}=\frac{u_{1}}{m_{1}}p_{xy} + \frac{u_{12}}{m_{1}}q_{xy} - \alpha_{yz} \alpha_{zx},  \\
\dot{\alpha}_{yz}=\frac{u_{1}}{m_{1}}p_{yz} + \frac{u_{12}}{m_{1}}q_{yz} - \alpha_{zx} \alpha_{xy},  \\
\dot{\alpha}_{zx}=\frac{u_{1}}{m_{1}}p_{zx} + \frac{u_{12}}{m_{1}}q_{zx} - \alpha_{xy} \alpha_{yz},  
\end{eqnarray}
\end{subequations}
and the similar forms by replacing all indices 1 by 2 and ($p,\alpha$) by ($q,\beta$), where $p^{(0)}_{x,y,z}$ and $q^{(0)}_{x,y,z}$ are the coefficients of the equilibrium condensate obtained by Eqs. (\ref{2compTFden}a) and (\ref{2compTFden}b).
These equations show that all scissors modes should initially be excited to induce the coupling between scissors modes. 
This conclusion was also obtained by Al Kawaja and Stoof for a single-component \cite{Kawaja}. 

\subsection{Numerical simulation} \label{numeridis}
To complement the analytical results obtained above, we study the dynamics of two-component BECs by the numerical simulations of the time-dependent GP equations (\ref{2tgpe}). 
For simplicity, we study the case of $a_{1}=a_{2}$ and $m_{1}=m_{2}$, and let $a_{12}$ and $\lambda=\omega_{y}/\omega_{x}$ be the variable parameters. 
We confine ourselves to the two-dimensional system, which is adequate for studying the mode coupling. 
A two-dimensional TF analysis of two-component BECs yields six normal modes; when $a_{12}=0$, these are a $xy$-scissors mode with the frequency $\Omega_{xy}^{2}/\omega_{x}^{2}=1+\lambda^{2}$ and two diagonal quadrupole modes with the frequency $\Omega_{\mp}^{2}/\omega_{x}^{2} = (3 + 3 \lambda^{2} \mp \sqrt{9 \lambda^{4} - 14 \lambda^{2} +9})/2$, corresponding to the $m=0$ low- and high-lying modes in the three-dimensional case. 
The frequencies of the out-of-phase modes are multiplied by a factor $\delta a$ given in Eq. (\ref{smatome}), decreasing with $a_{12}$.
The numerical calculations are done using an alternating direction implicit method as explained in Ref. \cite{Kasamatsu}. 
Initially, we search the equilibrium solution of two-component BECs by the norm-conserving imaginary time propagation of Eqs. (\ref{2tgpe}a) and (\ref{2tgpe}b). 
Starting from this initial state with appropriate modulation of the phase profile, we calculate the time evolution of the condensates. 
The parameter values are followed by the condition in Ref. \cite{Marago}; the number of the particles is $10^{4}$ and the aspect ratio of the trapping frequency is $\omega_{y}/\omega_{x}=\sqrt{8}$. 

We first numerically check the excitation frequencies obtained analytically in the TF approximation. In order to excite the collective modes, we change the phase $\theta_{i}(x,y)$ of the obtained initial state by a small factor $\epsilon_{i}=0.005$; for the $xy$-scissors mode $\theta_{i}(x,y) = \epsilon_{i} x y$, and for the $m=0$ low- and high-lying mode $\theta_{i}(x,y) = \epsilon_{i} (x^{2} + q_{\pm} y^{2})$ with 
$q_{\pm}=(-3 + 3 \lambda^{2} \pm \sqrt{9-14 \lambda^{2} + 9 \lambda^{4}})/2$ \cite{Dalfovo}. 
For the in-phase (out-of-phase) motion we set $\epsilon_{1}=(-) \epsilon_{2}$. 
We calculate the time evolution of the mean values $\langle x^{2} \rangle_{1,2}$, 
$\langle y^{2} \rangle_{1,2}$, and $\langle x y \rangle_{1,2}$, fitting a sinusoid to the calculated data in order to extract frequencies, where the first two values characterize the diagonal quadrupole modes and the last one does the scissors mode.
By varying intercomponent interaction $a_{12}$, we plot the frequencies of the three collective oscillations in Fig. \ref{ep=0m2=m1}. 
The numerical results agree fairly well with the analytical results, but as $a_{12}/a_{1}$ approaches unity, the frequencies of the out-of-phase motion shift upward from the analytical ones. 
This is associated with the large amplitude oscillation induced spontaneously because of the softening of the frequencies or the deviation from the approximation with the overlapping TF profile. 

\begin{figure}[btp]
\includegraphics[height=0.38\textheight]{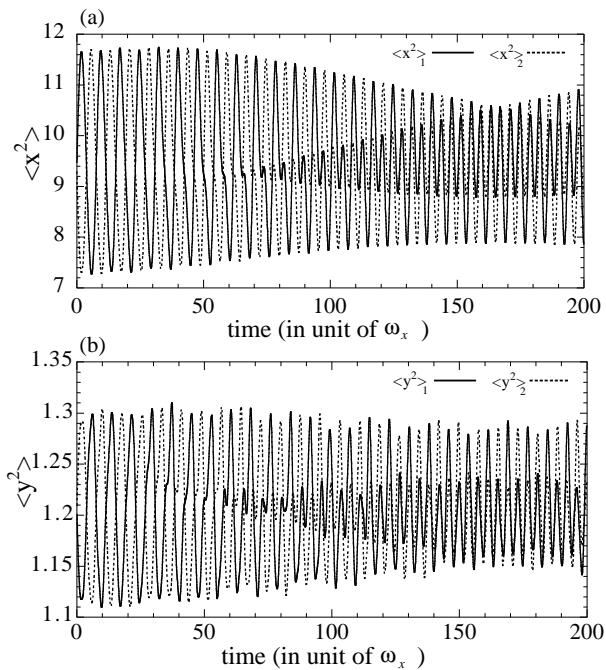}
\caption{Time development of (a) $\langle x^{2} \rangle_{1,2}$ and (b) $\langle y^{2} \rangle_{1,2}$, showing the mode-coupling process of (VI). 
The parameters used are $a_{12}/a_{1}=0.665$, $\lambda=\sqrt{8}$, and modulation factor $\epsilon=0.05$. } 
\label{dyna1m=0}
\end{figure}
\begin{figure}[btp]
\includegraphics[height=0.38\textheight]{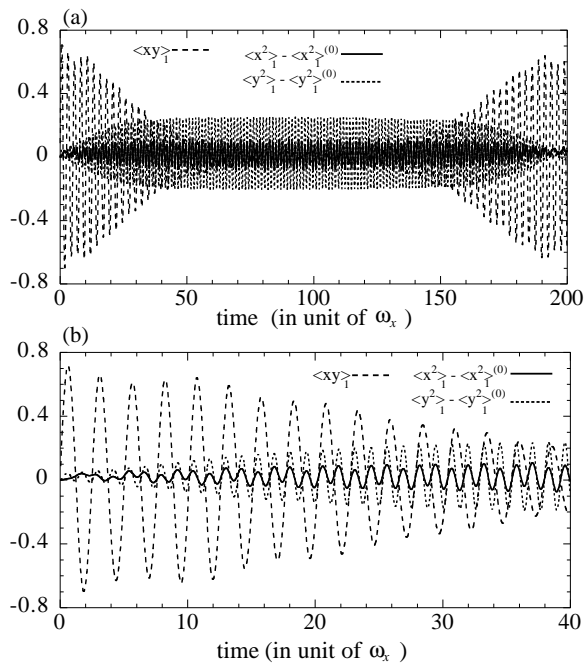}
\caption{(a) Time development of $\langle xy \rangle_{1}$, $\langle x^{2} \rangle_{1}-\langle x^{2} \rangle_{1}^{(0)}$ and $\langle y^{2} \rangle_{1}-\langle y^{2} \rangle_{1}^{(0)}$, where $\langle  \rangle^{(0)}$ represents the initial value, showing the mode-coupling process of (IX). 
The parameter values are $a_{12}/a_{11}=0.240$, $\lambda=\sqrt{8}$, and modulation factor $\epsilon=0.2$. (b) An enlarged view of (a) for the time interval [0,40].}
\label{dyna2sci}
\end{figure}

Next, we investigate the dynamics of the mode coupling when the second harmonic of a lower mode coincides with the frequency of a higher mode. 
Numerical simulation shows that the oscillation indeed exhibits significant nonlinear behavior because of the second harmonic generation. 
We show two examples of this nonlinear dynamics in Figs. \ref{dyna1m=0} and \ref{dyna2sci}. 
Figure \ref{dyna1m=0} shows the time development of the mean widths of the condensates for $\lambda=\sqrt{8}$ and $\delta a=0.20$ ($a_{12}/a_{1}=0.665$), when we excite initially the $m=0$ low-lying out-of-phase mode. 
The condensate widths $\langle x^{2} \rangle_{1,2}$ and $\langle y^{2} \rangle_{1,2}$ undergo an out-of-phase oscillation, and $\langle x^{2} \rangle_{1}$ ($\langle y^{2} \rangle_{1}$) and $\langle x^{2} \rangle_{2}$ ($\langle y^{2} \rangle_{2}$) undergo an out-of-phase oscillation, too. 
As time passes, keeping $\langle x^{2} \rangle_{1,2}$ and $\langle y^{2} \rangle_{1,2}$ out-of-phase, the oscillations of $\langle x^{2} \rangle_{1}$ and  $\langle x^{2} \rangle_{2}$ as well as $\langle y^{2} \rangle_{1}$ and $\langle y^{2} \rangle_{2}$ change from out-of-phase to in-phase as shown in Fig. \ref{dyna1m=0}. 
This time development shows directly the process of (VI) in Table \ref{tab:table2}. 
When we calculate under the condition $\delta a=0.25$ ($a_{12}/a_{1}=0.60$) shown in Table \ref{tab:table2} (VI), this nonlinear behavior is weaker than that in Fig. \ref{dyna1m=0}. 
The deviation of the resonance from $\delta a=0.25$ to $\delta a =0.2$ is due to the frequency shift appearing in the present numerical simulation. 
After a long time, the oscillation gets back to the original out-of-phase behavior. 
This is nothing but the recurrence phenomena known as the Fermi-Pasta-Ulam problem \cite{Fermi}; the similar behavior is also found in the dynamics of a rotating BEC \cite{Kasamatsu}. 

The mode coupling from the out-of-phase $xy$-scissors mode to the $m=0$ high-lying in-phase mode ((IX) in Table \ref{tab:table2}) is shown in Fig. \ref{dyna2sci}. 
Figure \ref{dyna2sci}(a) shows the time development of the mean values for component 1 for $\lambda=\sqrt{8}$ and $\delta a = 0.613$ ($a_{12}/a_{1}=0.24$). 
From an enlargement view shown in Fig. \ref{dyna2sci}(b), we see that the oscillation of the mean value $\langle xy \rangle_{1}$ is damped quickly, converting into the oscillations of the diagonal component $\langle x^{2} \rangle_{1}$ and $\langle y^{2} \rangle_{1}$, where they undergo an in-phase oscillation; those of the component 2 oscillate in the same manner. 
Again, the oscillation recovers the out-of-phase of the $xy$-scissors mode. 

\section{SUMMARY AND CONCLUSION}\label{concl}
In summary, we have studied the diagonal quadrupole modes and the scissors modes of two-component BECs, and the nonlinear dynamics associated with the second harmonic generation of the quadrupole and scissors modes. 
Within the TF approximation and quadratic polynomial ansatz for the density fluctuation, the linear analysis leads to analytical expressions for the collective-mode frequencies. 
These analytical results are confirmed by the numerical simulation of the GP equations. 
Including the nonlinear contribution of the fluctuation, we have clarified a variety of the mode couplings between collective modes. 
The condition of the second harmonic generation is obtained as functions of the aspect ratio of a trapping potential and intercomponent interaction strength. 
At the resonance of two collective modes, the collective dynamics of two-component condensates shows remarkable nonlinear behavior involving damping and recurrence of the density oscillation. 

In this paper, we have confined ourselves to the condensates with their density profiles completely overlapping. 
Two-component BECs consisting of the same atoms realized so far are mixtures of the states $|1,-1 \rangle$ and $|2,1 \rangle$ of $^{87}$Rb in JILA \cite{Hall}, $|2,1 \rangle$ and $|2,2 \rangle$ of $^{87}$Rb in LENS \cite{Maddaloni} and $|1,0 \rangle$ and $|1,1 \rangle$ of $^{23}$Na in MIT \cite{Stenger}. 
Although the scattering lengths of the mixture in JILA satisfy the overlapping condition, the resulting double condensates do not form the inverted parabolic density profile. 
The latter two mixtures lie slightly in the phase-separated region. 
Moreover, a mixture of $^{41}$K and $^{87}$Rb BECs also lies deeply in a phase-separate region \cite{Modugno}. 
By changing the scattering lengths via the Feshbach resonance, however, we can, in principle, prepare the condensates in an overlapping regime and verify our prediction \cite{Burke}. 
We believe that this work will help the understanding of the complex dynamical behavior of mutually interacting multicomponent BECs.

\begin{acknowledgements}
K.K. and M.T. acknowledge support by a Grant-in-Aid for Scientific Research (Grant No.1505955 and 15340122, respectively)
by the Japan Society for the Promotion of Science.
M.U. acknowledges support by a Grant-in-Aid for 
Scientific Research (Grant No.14740181) by 
the Ministry of Education, Culture, Sports, Science and Technology 
of Japan, and a CREST program by Japan Science and Technology 
Corporation (JST). 
\end{acknowledgements}



\begin{thebibliography}{99}
\bibitem{tyuu}
The relative displacement of the potential minima can be tuned or vanished by adjusting the external parameter, as explained in Ref. \cite{Hall}.
\end{thebibliography}
\end{document}